\documentclass[aps,prl,amsmath,twocolumn]{revtex4}
\usepackage{bm}

\begin{document}

\title{Comment on the paper 'Saturation and Multifractality of Lagrangian and Eulerian Scaling Exponents
in Three-Dimensional Turbulence' }

\author{ V.A. Sirota,
\footnote{\\ * Electronic address: sirota@lpi.ru}  K.P. Zybin }

\affiliation{119991 P.N.Lebedev Physical Institute, Russian Academy of Sciences, Moscow, Russia}


\begin{abstract}

\end{abstract}

\maketitle
The paper \cite{Sreenivasan} brings long-awaited clarity into the complicated picture of the latest
numerical and experimental results and sets the 'state-of-the-art' in the problem of Lagrangian and
Eulerian velocity scaling. In particular, in \cite{Sreenivasan} it is convincingly shown that the
scaling exponents of transverse Eulerian velocity structure functions saturate, and the saturated
value is $\zeta_{\perp \infty}\simeq 2.1$.
  It would be interesting  and useful to
 understand the physical issue of this saturation.
  What does this statistical phenomenon mean  physically?  The aim of this comment is to
  draw attention to 
   the objects that may be responsible for this behavior of the structure functions.

It has been known long ago (see, e.g.,\cite{Farge}) that the high-vorticity regions make crucial
contribution to the intermittent velocity statistics. The conception of vortex filaments acting as
a source of intermittency  explains  naturally the predominance of the transverse velocity
differences over the longitudinal ones.

The saturation of the scaling exponents can also be understood in the frame of this approach.
 From dynamical point of view, the main process in the three-dimensional turbulence is the
 stretching of vortex filaments. It is responsible for production of filaments and, thus,  for
 (at least  transverse) scaling; the saturation
 means that this process stops at some limit. What is the limit?

 In \cite{arXiv, UFN} we showed that the stretching is restricted by a logarithmic singularity
 of pressure: the critical, 'extreme' vortex has flat velocity profile and demonstrates singular
 behavior of velocity in the center,
 \begin{equation}\label{extrim}
{\bf v}(z,\rho,\phi) = c {\bf e}_{\phi}
\end{equation}
Actually, let a vortex filament have velocity distribution $v\propto \rho^{\gamma}$. The smaller
$\gamma$, the stronger the singularity along the axis, the more significant the contribution of the
filament to the structure functions of large orders $p$. However, since pressure and velocity are
finite, $\gamma$ cannot be negative; the extreme filament (\ref{extrim})  is just what corresponds
to $\gamma=0$.
We calculated the contributions to the transverse and longitudinal structure functions produced by
this critical vortex for large $p$; 
these are
\begin{equation}
\label{extrim2}
  \langle \left| \Delta v \times {\bf l}/l \right| ^p  \rangle \propto
  \frac{2^p}{\sqrt{p}} \, l^2\,,\quad
  \langle \left| \Delta v \cdot {\bf l}/l \right| ^p \rangle \propto
   p^{-5/2} l^2
\end{equation}
The scaling exponent is 2 in both cases. However, as $p$ increases, the contribution of the extreme
filament to the transverse structure function increases and becomes determining, while its
contribution to the longitudinal structure function decreases to zero. This is natural, since
longitudinal velocity increments in vortices are evidently small compared to the transverse
increments. So, the straight extreme filament (\ref{extrim}) does not contribute to the
longitudinal structure functions; but the bent filaments do.  The curved sections of filaments
contribute to the longitudinal structure function. However, the account of the curvature results in
the next order in $l/L$: $\zeta_{|| \infty} \simeq 3$.

Eventually, the prediction of \cite{arXiv, UFN} was:
$$
\zeta_{\perp \infty} \simeq 2  \ , \quad p \ge 10 \ , \quad \zeta_{\|  \infty} \simeq 3  \ , \quad
p \ge 16
$$
 The prediction for transverse scaling exponents coincides with the results of the paper
\cite{Sreenivasan}.
 So, the hypothesis of existence of long-living objects with the velocity profiles close to the
 critical one provides  a simple physical picture that, on one hand, predicts correct values for
 the transverse
  scaling exponents, including their saturation, and, on the other, naturally explains the difference
  between longitudinal and
 transverse scalings. It is interesting whether it will be able to predict the
 saturation of longitudinal scaling exponents as well.

\end{document}